\documentclass[10pt,twocolumn,letterpaper]{article}

\usepackage{cvpr} 
\usepackage{times}
\usepackage{epsfig}
\usepackage{graphicx}
\usepackage{amsmath}
\usepackage{amssymb}
\usepackage[numbers,sort&compress]{natbib}
\usepackage{tabularray}
\usepackage{adjustbox}
\usepackage{amsmath}
\usepackage{graphicx}
\usepackage{placeins}
\usepackage{doi}
\usepackage{hyperref}
\usepackage{booktabs, multirow}

\usepackage{caption}
\begin{document}

\title{Quantitative and Qualitative Comparison of Generative Models for Subject-Specific Gaze Synthesis: Diffusion vs GANs}

\author{
Kamrul Hasan, Dmytro Katrychuk, Mehedi Hasan Raju, Oleg V. Komogortsev\\
Texas State University, San Marcos, Texas, USA\\
{\tt\small \{kamrul.hasan, d\_k139, m.raju, ok\}@txstate.edu}
}

\maketitle
\thispagestyle{empty}

\begin{abstract}
Gaze-based biometrics has emerged as a promising approach for user authentication, but advances in this area are constrained by the limited availability of high-quality, subject-specific gaze recordings. Recent generative models have shown promise for synthesizing gaze data, yet most existing approaches rely on random noise distributions or global, predefined latent embeddings and do not explicitly model subject-specific gaze characteristics. To address this limitation, we revisit two recent generative models, diffusion and generative adversarial networks (GANs), and modify both to support subject-aware gaze synthesis. For the diffusion-based approach, we incorporate compact user embeddings to capture subject-level gaze traits. For the GAN-based approach, we introduce a subject-specific conditioning module that guides the generator to preserve idiosyncratic gaze patterns. Later, we evaluate both approaches using standard eye-movement signal quality metrics, including spatial accuracy and precision, and assess whether the generated sequences retain identity-related features relevant to biometric applications. Experimental results show that the diffusion-based approach produces more realistic, identity-preserving gaze sequences than the GAN-based approach. Overall, this work advances the understanding of synthetic gaze quality, realism, and subject specificity and supports the development of gaze-based biometric applications. 
\end{abstract}


\section{Introduction}
Gaze-based authentication is an important approach for enabling secure authentication on eXtended Reality (XR) devices \cite{lohr2024establishing,lohr2026gaze}, especially those that employ non-image-based sensors \cite{aziz2022synchroneyes}. Beyond authentication, gaze analysis has also shown promise in healthcare, where eye movements can support dyslexia identification \cite{haller2022eye} and autism spectrum disorder (ASD) detection \cite{thanarajan2023eye}, as well as in security applications such as liveness detection \cite{komogortsev2015person, raju2022iris} based on distinctive eye-movement dynamics.
These applications and developments position eye movement as a unique and informative behavioral signal for biometric identification, user authentication, and privacy-aware human–computer interaction (HCI).

However, to make these applications a reality, high-sampling-rate (e.g., 250–1000 Hz) eye movement signals are crucial, as they capture micro-oculomotor events (such as microsaccades and fine-grained smooth pursuit) and extract subject-specific patterns that are less precise at lower sample rates (e.g., 30–60 Hz). 
In addition, high-frequency gaze sequences have detailed subject-specific information that can enable re-identification \cite{khan2019survey} and the inference of sensitive attributes (e.g., gender, identity, ethnicity), raising significant privacy concerns \cite{steil2019privacy}. 
Despite the value of such signals for authentication and other applications, most large datasets are proprietary and publicly unavailable \cite{garbin2020dataset, cheng2024appearance}. 
Moreover, collecting large-scale, diverse, and high-quality datasets is expensive and labor-intensive, requiring specialized equipment, controlled acquisition protocols, long recording sessions, and expert annotation of oculomotor events \cite{griffith2021gazebase}.
Together, these challenges motivate us to address these gaps by focusing on subject-aware synthetic data generation that should preserve necessary properties while reducing reliance on real data.

Synthetic data generation is emerging as a promising solution to address data shortages and mitigate privacy concerns \cite{liu2025preserving}.
More recently, with the wider adoption of generative adversarial networks (GANs) \cite{goodfellow2020generative}, SP-EyeGAN \cite{prasse2023sp} generates realistic temporal scanpaths, including both fixations (periods of relatively stable gaze) and saccades (rapid eye movements between fixation points).
However, they use a random noise distribution to generate signals that are not specific to the target, which limits the approach's practicality in real applications.
More recently, denoising diffusion probabilistic models (DDPMs) \cite{ho2020denoising} have emerged as a powerful framework for time-series generation. For example, DiffEyeSyn \cite{jiao2024diffeyesyn} conditions the diffusion process on identity-removed signals and user embeddings to incorporate user-specific attributes while preserving natural gaze trajectories.
These methods motivate us to conduct a systematic experiment of how to encode subject identity and which generative approach most effectively preserves individual gaze behavior.

Building on these foundations, we instantiate subject-specific variants of both generative approaches. 
In terms of diffusion, we adhere to DiffEyeSyn's \cite{jiao2024diffeyesyn} approach, introducing several conditioning and training modifications.
First, we derive the conditional signal by downsampling to 25 Hz and then upsampling to the raw rate (i.e., 1000 Hz), motivated by evidence that subject identity above low frequencies can be removed without discarding task-relevant kinematics \cite{raju2023determining}. 
Secondly, we substitute the 512-dimensional conditioning vector with a concise 128-dimensional user embedding. 
Collectively, these modifications yield more authentic, subject-specific synthetic gaze signals. 
Similarly, in the GAN-based approach, we enhanced SP-EyeGAN's \cite{prasse2023sp} dual generators for fixations and saccades by incorporating a Subject-Specific Condition Generator (SCG) module with two branches, namely the Directional Data Quality Feature Extractor (DDQFE) and the one-hot (OH) encoder. 
This quality-aware conditioning improves authenticity and subject specificity while maintaining a lightweight structure.

Finally, we conduct a rigorous head-to-head evaluation of the revised diffusion- and GAN-based approaches under the same experimental setting. We report standard signal-quality metrics, including spatial accuracy, spatial precision, and cosine similarity between embeddings derived from synthetic and real gaze signals. We also evaluate biometric utility by examining whether the generated gaze sequences preserve identity-discriminative dynamics for authentication and identification.
Overall, the specific contributions of our work are twofold: 
\begin{enumerate}
\item A compact-embedding and identity-removed signal conditioning variant of DiffEyeSyn that improves subject specificity with lower complexity.
Additionally, an enhanced SP-EyeGAN with an explicit Subject-Specific Condition Generator that fuses DDQFE's features with one-hot encoding subject identity to guide realistic fixation and saccade generation.
\item A qualitative and quantitative comparison of the above-mentioned approaches for subject-specific gaze synthesis, as well as how they affect biometric user-authentication performance.
\end{enumerate}

\section{Related Work}

\subsection{Statistical Models}
Earlier models for eye movement synthesis relied primarily on training-free statistical methods grounded in signal processing \cite{lee2002eyes, duchowski2015modeling, duchowski2016eye}.
As such, approaches ranging from saccade-based animation \cite{lee2002eyes} and head-gaze coordination \cite{ma2009natural} to photorealistic image generation. For example, SynthesEyes \cite{wood2015rendering} was used to model realistic eye movements and gaze behavior.
While these studies established foundational realism, they did not capture the full complexity of human oculomotor behavior.

Subsequent statistical models \cite{le2012live, fuhl2018eye_Kasneci, fuhl2018eye_Santini, yeo2012eyecatch, lan2022eyesyn} advanced gaze synthesis by integrating known oculomotor characteristics. 
EyeCatch \cite{yeo2012eyecatch} used a Kalman filter \cite{welch1995introduction} to simulate fixations, saccades, and smooth pursuits. 
EyeSyn \cite{lan2022eyesyn} further introduced a physics-based framework that models these movements with psychology-inspired equations and adds realistic jitter through Gaussian drift and pink noise. 
Despite these advances, training-free methods still yield average behavior, overlook individual variability, and simplify eye movements, such as microsaccades and tremors.

\subsection{Machine-learning Models}

To address the limitations of statistical models, researchers have recently concentrated on machine learning approaches, particularly deep neural networks \cite{simon2016automatic, assens2018pathgan, goodfellow2020generative, kaur2020eyegan, jiao2023supreyes}. 
Initial research integrated visual inputs with sequence models to predict human-like scanpaths.
For example, Assens et al. \cite{assens2018pathgan} proposed PathGAN, which comprises GANs that output likely fixation points on an image. 
PathGAN could replicate where individuals looked but failed to simulate continuous eye movements; therefore, it could not capture actual saccadic trajectories or velocity profiles.

In addition to image-conditioned models, multiple research \cite{fuhl2021fully, de2022next, fuhl2022hpcgen} investigated stimulus-agnostic approaches to synthesizing eye movement data.  Fuhl et al. \cite{fuhl2021fully} utilized fully convolutional networks for semantic segmentation, reconstruction, and variational autoencoder (VAE)-based generation of raw eye-tracking data without preprocessing. 
As it was not conditioned on user identity or task, the generated sequence represented generic behavior averaged over many individuals.
Therefore, recent studies have begun incorporating adversarial \cite{goodfellow2020generative} and diffusion models \cite{ho2020denoising, nichol2021improved} to synthesize eye movement signals. 
SP-EyeGAN \cite{prasse2023sp} introduced a GAN-based approach explicitly designed to produce fixations and saccades. 
DiffGaze \cite{jiao2025diffgaze}, a diffusion-based model was introduced to synthesize scanpaths at 30 Hz for 360\textdegree VR images. 
However, both SP-EyeGAN and DiffGaze generate stimulus-driven eye-movement behavior without explicit subject-identity conditioning.
Meanwhile,  Jiao et al. \cite{jiao2024diffeyesyn} proposed DiffEyeSyn, the first approach to generate user-specific eye movements. 

\section{Implemented Architectures}
\begin{figure*}[htbp]
    \centering
    \includegraphics[width=0.8\linewidth]{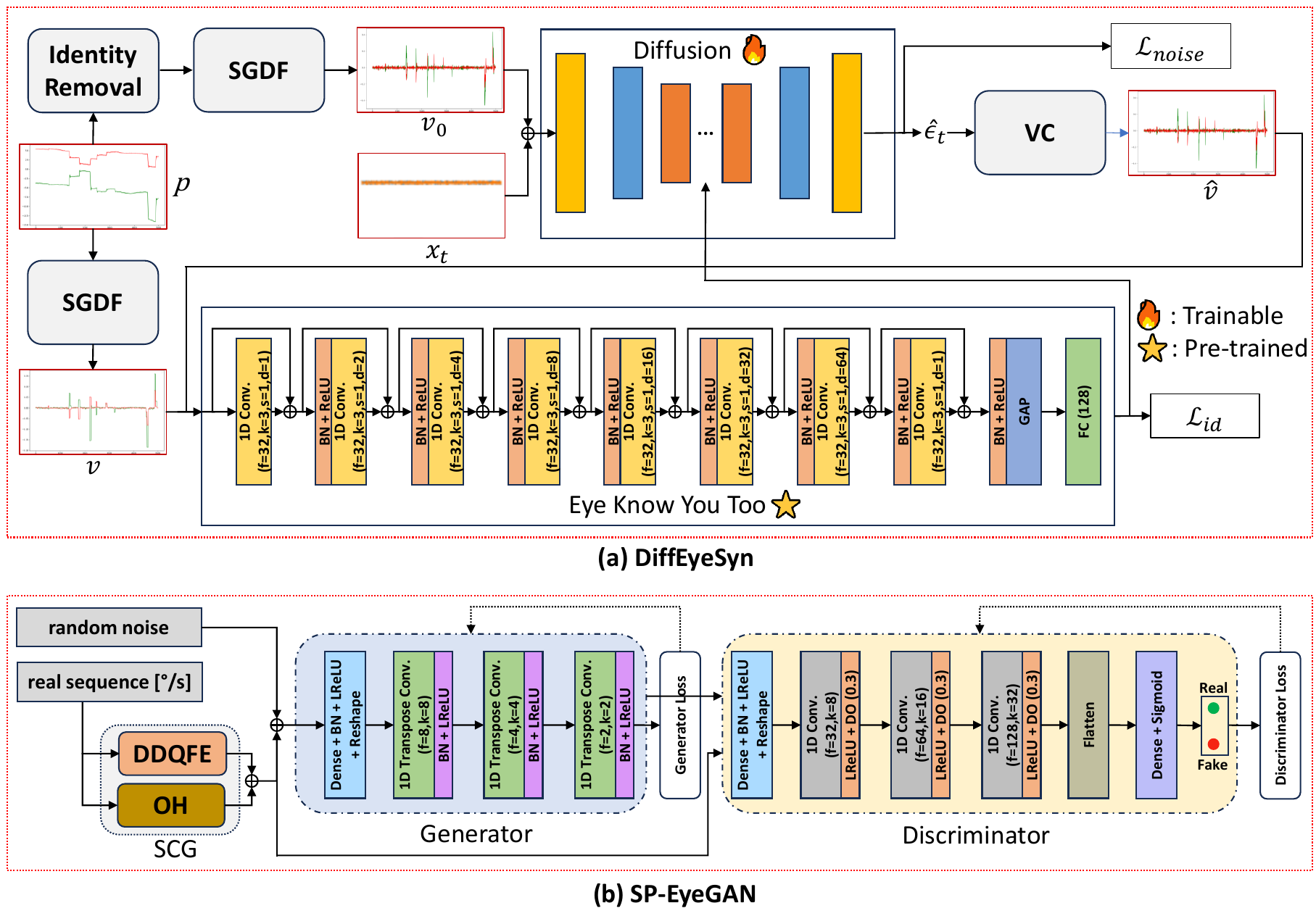}
    \caption{(a) Overview of the updated DiffEyeSyn architecture: $p$, $v_0$, $v$ denote real positional, identity-removed, and real velocity signal, respectively. At each diffusion step $t$, noise $x_t$ and $v_0$ serve as input to the diffusion model. 
    (b) Overview of the updated SP-EyeGAN architecture comprises Subject-Specific Condition Generator (SCG), Generator, and Discriminator.
}
    \label{fig:main_architecture}
\end{figure*}

The overall architectures for both generative approaches are shown in Figure \ref{fig:main_architecture}(a) (i.e., DiffEyeSyn) and Figure \ref{fig:main_architecture}(b) (i.e., SP-EyeGAN).

\subsection{DiffEyeSyn}

\subsubsection{Data Preprocessing}
Let the gaze sequence be $p \in \mathbb{R}^{S \times d}$, where $S$ is the number of samples in milliseconds $(ms)$ and $d=2$ denotes horizontal and vertical positions. As illustrated in Figure \ref{fig:main_architecture}(a), $p$ is first fed into the Identity Removal module. 
Then, building on prior work \cite{raju2023determining} and our analysis, $p$ is downsampled to 25 Hz and subsequently upsampled back to 1,000 Hz. 
Later on, the identity-removed positional signal ($^{\circ}$) and raw positional signal ($^{\circ}$) are converted into identity-removed velocity ($v_0$) and raw velocity signal ($v$) using the Savitzky-Golay \cite{savitzky1964smoothing} differentiation filter (SGDF). 
Subsequently, invalid velocity samples (e.g., NaNs) are assigned a value of 0. 
Later, velocities are constrained to the interval [-1000, 1000] °/s, then rescaled to the range [-90, 90] and normalized to [-1, 1] using a sine-based transformation that bounds extreme amplitudes while preserving relative dynamics.

\subsubsection{Conditional Diffusion Model}
For identity-removed velocity signals $v_0$, in the diffusion step \(t\), the noised sample is formed as follows:

\begin{equation}
x_t=\sqrt{\bar\alpha_t}\, v_0+\sqrt{1-\bar\alpha_t}\,\varepsilon_t,\quad 
\varepsilon_t\sim\mathcal{N}(0,I)
\end{equation}

\noindent where $x_t$ is the noisy sample at step $t$, $\bar{\alpha}_t$ is the cumulative noise-retention factor, $\varepsilon_t$ is the Gaussian noise, $\mathcal{N}(0, I)$ is the normal distribution, and $t\in\{1,\dots,T\}$ denotes the number of discrete diffusion time steps (i.e., the length of the forward noising and reverse denoising Markov chain).

To guide subject-specific generation, we use a pre-trained Eye Know You Too (EKYT) encoder \cite{lohr2022ekyt}, denoted by $\phi(\cdot)$, to extract a user embedding $z=\phi(v)$. Instead of the original 512-D EKYT embedding, we use a compact 128-D embedding learned from a single cross-validation fold. The noise prediction is performed as follows:

\begin{equation}
    \begin{split}
\hat\varepsilon_t=\varepsilon_\theta(x_t,\,t,\,\text{cond}),\qquad \text{cond}=(v_0,\,z),\; z=\phi(v)
    \end{split}
\end{equation}

\noindent where $\varepsilon_\theta(\cdot)$ is the denoiser parameterized by $\theta$, and $\text{cond}=(v_0,z)$ combines the identity-removed signal with the EKYT-derived user embedding. We inject this conditioning through FiLM-based residual modulation, where the conditioning features generate feature-wise scale and shift parameters for intermediate denoiser representations. The predicted noise $\hat{\varepsilon}_t$ is then used by the velocity-converter (VC) module to reconstruct the velocity signal:
\begin{equation}
\hat v=\frac{x_t-\sqrt{1-\bar\alpha_t}\,\hat\varepsilon_t}{\sqrt{\bar\alpha_t}}
\end{equation}

The training objective includes a noise-prediction loss $\mathcal{L}_{\text{noise}}$ for reconstructing the injected Gaussian noise and an identity-guidance loss $\mathcal{L}_{\text{id}}$ that aligns the embedding of the generated velocity $\hat{v}$ with that of the real velocity $v$.

\subsection{SP-EyeGAN}

\subsubsection{Data Preprocessing}
Following SP-EyeGAN \cite{prasse2023sp}, we train two GANs, FixGAN and SacGAN, for fixation and saccade generation, respectively. Raw 1,000 Hz gaze signals are first segmented into fixations and saccades using the Dispersion--Threshold Identification (I-DT) algorithm \cite{salvucci2000identifying}. For each segment, positional samples are smoothed and converted to 2D velocities ($^{\circ}/s$) using SGDF \cite{savitzky1964smoothing}. Invalid velocity samples (e.g., NaNs) are replaced with 0, and the resulting fixation and saccade segments are used to train FixGAN and SacGAN, respectively.

\subsubsection{Subject-Specific Condition Generator}
The Subject-Specific Condition Generator (SCG) module comprises two components: the Directional Data Quality Feature Extractor (DDQFE) and the one-hot (OH) encoder. The DDQFE incorporates horizontal and vertical velocities into displacement profiles and derives two metrics per axis: spatial dispersion (i.e., standard deviation) and temporal energy (i.e., root-mean-square error). These features summarize segment-level data quality and movement intensity. Furthermore, subject identity is represented as an OH representation with a dimensionality equal to the number of training subjects. Thereafter, the DDQFE features and OH identity vector are concatenated to form the final conditioning vector.

\subsubsection{Conditional GAN Model}
The extended conditional architecture of the GANs is illustrated in Figure \ref{fig:main_architecture}(b).
For a given sample from the fixation\_dataset $f_v \in \mathbb{R}^{T \times d}$, where $T$ is the number of time steps in $ms$ and $d=2$ corresponds to the horizontal and vertical velocity components (°/s).
Let $c$ denote the SCG conditioning vector, $z$ the latent noise, $G(z,c)$ the generated velocity sequence, and $D(\cdot)$ the discriminator output. The conditional adversarial losses are:

\begin{equation}
\begin{aligned}
\mathcal{L}_{D}
&= \mathbb{E}_{(f_v,c)\sim p_{\text{data}}}
   \!\left[-\log D(f_v,c)\right] \\
&\quad + \mathbb{E}_{z\sim p(z),\,c}
   \!\left[-\log\!\big(1 - D(G(z,c),c)\big)\right]
\end{aligned}
\end{equation}

\begin{equation}
\mathcal{L}_{G}
= \mathbb{E}_{z\sim p(z),\,c}\!\left[-\log D\!\big(G(z,c),c\big)\right]
\end{equation}

\noindent where $\mathcal{L}_{D}$ and $\mathcal{L}_{G}$ denote the discriminator and generator binary cross-entropy losses, $p_{\text{data}}$ is the real distribution over $(f_v,c)$, and $p(z)$ is the latent noise prior.

Finally, after training FixGAN and SacGAN, we create complete gaze trajectories using fixation and saccade sequences. We then identify $n$ fixation points before building the trajectory by concatenating the fixation segments with saccades. For additional details, please refer to the SP-EyeGAN \cite{prasse2023sp} paper.

\section{Experiments}
\subsection{Dataset}
We used the publicly available GazeBase dataset \cite{griffith2021gazebase} for training and evaluation. GazeBase was collected with an EyeLink 1000 eye tracker and contains monocular left-eye recordings sampled at 1,000 Hz from 322 participants across nine rounds over 37 months, totaling 12,334 recordings. It includes seven tasks: fixation (FXS), horizontal saccade (HSS), random oblique saccade (RAN), reading (TEX), free-viewing video tasks (VD1 and VD2), and gaze-driven gaming (BLG). Among these, only HSS and RAN provide corresponding stimulus signals, which are required for spatial accuracy and precision evaluation.
To ensure compatibility with the pre-trained EKYT model \cite{lohr2022ekyt}, we adopted a similar split across the six training/evaluation tasks, excluding BLG from model training. The training set contains 263 participants and 63,161 non-overlapping 5-second sequences sampled at 1,000 Hz. The test set contains 59 unseen participants from round 6. During evaluation, synthetic 5-second sequences are concatenated in temporal order to match the corresponding real recording length, enabling direct synthetic-real comparison. No subjects overlap between training and testing, ensuring evaluation on completely unseen identities.

\subsection{Implementation Details}
We trained DiffEyeSyn on 5s windows sampled at 1,000 Hz (5,000 samples per window) using a diffusion process with $T=50$ and a linear scheduler from 0.0001 to 0.05, where $T$ is the number of discrete diffusion time steps in the forward and reverse processes. Additionally, AdamW optimizer \cite{loshchilov2017decoupled} was utilized with a learning rate of 0.0002 and a batch size of 32, and the model was trained for 450 epochs. Meanwhile, SP-EyeGAN was trained for 100 epochs with a batch size of 32 and a learning rate of 0.0001. Since the original SP-EyeGAN model was trained only on the TEX stimulus, we trained all tasks separately, as FixGAN/SacGAN generate fixations and saccades rather than the entire signal. All experiments were implemented using Python 3.9.12, CUDA 12.4, PyTorch 2.1.0, PyTorch-Lightning 1.9.5, and TensorFlow 2.14.0 and were run on a single NVIDIA RTX A6000 GPU (48 GB GDDR6). Our source code and trained models are available on the Texas State Digital Collections Repository at \url{https://hdl.handle.net/10877/25137}.

\section{Results}
We evaluate synthetic gaze quality using spatial accuracy and spatial precision, reported at both the error percentile (E) and user percentile (U) levels \cite{aziz2024evaluation}. For these metrics, we consider stable fixation periods by identifying 80-\textit{ms} fixation bins for evaluation \cite{lohr2019evaluating}. We further report synthetic-real embedding similarity, biometric authentication performance, and an ablation study. As the central objective of this work is a controlled diffusion-versus-GANs comparison, the variational autoencoder (VAE) is included only as an auxiliary baseline for embedding similarity and biometric evaluations, rather than as a full third model in spatial quality, qualitative, or ablation analyses.

\subsection{Spatial Accuracy}
Spatial accuracy was evaluated as the gaze-point error, measured in degrees of visual angle (dva), between the synthetic gaze signal and the target stimulus position. Table \ref{tab:spatial_accuracy} reports the spatial accuracy of the ground truth (real gaze signal), SP-EyeGAN, and DiffEyeSyn at both the user (U) and error (E) percentiles. Across both tasks, RAN and HSS, DiffEyeSyn consistently outperforms SP-EyeGAN. For example, at the median user level (U50), DiffEyeSyn achieves lower median error (E50), with 3.73 dva for HSS and 4.06 dva for RAN, whereas SP-EyeGAN shows substantially higher values, particularly for HSS (15.45 dva) and RAN (13.63 dva). Although neither model matches the ground truth, the lower errors indicate that DiffEyeSyn produces signals that are closer to the original gaze data and less affected by extreme outliers. This trend is also maintained under more challenging conditions, where DiffEyeSyn achieves lower error than SP-EyeGAN at higher user and error percentiles, such as (U95\textbar E95) for RAN (26.77 vs. 47.07 dva). Overall, these results suggest that DiffEyeSyn provides more accurate and consistent synthetic gaze signals across a broader range of users.

\begin{table}[h]
\centering
\caption{U\textbar{}E spatial accuracy for HSS and RAN. Ground truth denotes the real positional signal; $\downarrow$ indicates lower is better.}
\label{tab:spatial_accuracy}
\resizebox{0.6\columnwidth}{!}{%
\begin{tabular}{lcccc}
\toprule
\multirow{2}{*}{Model} & \multicolumn{2}{c}{U50\textbar{}E50 $\downarrow$} & \multicolumn{2}{c}{U95\textbar{}E95 $\downarrow$} \\
\cmidrule(lr){2-3} \cmidrule(lr){4-5}
 & HSS & RAN & HSS & RAN \\
\midrule
Ground truth & 0.89  & 0.92  & 30.85 & 21.80 \\
SP-EyeGAN    & 15.45 & 13.63 & 52.73 & 47.07 \\
DiffEyeSyn   & 3.73  & 4.06  & 38.61 & 26.77 \\
\bottomrule
\end{tabular}}
\end{table}

\subsection{Spatial Precision}
Spatial precision reflects the stability of the gaze signal and is commonly measured as the root mean square (RMS) dispersion of gaze points during steady fixations. We report precision in degrees RMS, where lower values indicate less signal jitter while the user is looking at a fixed target. Table \ref{tab:spatial_precision} presents the spatial precision of the ground truth (real gaze signal), SP-EyeGAN, and DiffEyeSyn. On average, both approaches achieve similar high precision. For example, at the median user level (U50), both SP-EyeGAN and DiffEyeSyn obtain an E50 precision of 0.01 degrees in HSS, indicating stable synthetic gaze during fixation. In the more challenging RAN task, both models show slightly higher jitter, likely due to small fixation instabilities between saccades. However, the difference becomes more apparent at higher user and error percentiles. Under worst-case conditions, SP-EyeGAN exhibits substantially larger precision errors than DiffEyeSyn. For instance, at U95\textbar E95 in the RAN task, SP-EyeGAN reaches 0.41 degrees RMS, whereas DiffEyeSyn remains lower at 0.29 degrees RMS. This suggests that some SP-EyeGAN-generated sequences contain stronger jitter or noise spikes, while DiffEyeSyn better limits such extreme outliers.

\begin{table}[h]
\centering
\caption{U\textbar{}E spatial precision for HSS and RAN. Ground truth denotes the real positional signal; $\downarrow$ indicates lower is better.}
\label{tab:spatial_precision}
\resizebox{0.6\columnwidth}{!}{%
\begin{tabular}{lcccc}
\toprule
\multirow{2}{*}{Model} & \multicolumn{2}{c}{U50\textbar{}E50 $\downarrow$} & \multicolumn{2}{c}{U95\textbar{}E95 $\downarrow$} \\
\cmidrule(lr){2-3} \cmidrule(lr){4-5}
 & HSS & RAN & HSS & RAN \\
\midrule
Ground truth & 0.01 & 0.01 & 0.86 & 1.14 \\
SP-EyeGAN    & 0.01 & 0.01 & 0.37 & 0.41 \\
DiffEyeSyn   & 0.01 & 0.01 & 0.35 & 0.29 \\
\bottomrule
\end{tabular}}
\end{table}

\begin{table}[h]
\centering
\caption{Cosine similarity between pre-trained EKYT embeddings of real and synthetic eye movement signals for the evaluated generative models. Here, `*' denotes the baseline version of that approach.}
\label{tab:cosine_similarity}
\resizebox{\columnwidth}{!}{%
\begin{tabular}{lccccc}
\toprule
\multirow{2}{*}{Task} & \multicolumn{5}{c}{Model} \\
\cmidrule(lr){2-6}
 & SP-EyeGAN* & SP-EyeGAN & VAE & DiffEyeSyn* & DiffEyeSyn \\
\midrule
BLG & $0.14 \pm 0.13$ & $0.16 \pm 0.12$ & $0.78 \pm 0.04$ & $0.92 \pm 0.04$ & $0.96 \pm 0.06$ \\
FXS & $0.11 \pm 0.15$ & $0.30 \pm 0.08$ & $0.64 \pm 0.06$ & $0.86 \pm 0.05$ & $0.92 \pm 0.08$ \\
HSS & $0.11 \pm 0.14$ & $0.24 \pm 0.11$ & $0.73 \pm 0.02$ & $0.91 \pm 0.03$ & $0.95 \pm 0.05$ \\
RAN & $0.13 \pm 0.14$ & $0.22 \pm 0.10$ & $0.71 \pm 0.03$ & $0.89 \pm 0.04$ & $0.95 \pm 0.04$ \\
TEX & $0.16 \pm 0.14$ & $0.21 \pm 0.10$ & $0.73 \pm 0.03$ & $0.89 \pm 0.03$ & $0.94 \pm 0.06$ \\
VD1 & $0.13 \pm 0.15$ & $0.31 \pm 0.10$ & $0.73 \pm 0.04$ & $0.90 \pm 0.04$ & $0.95 \pm 0.05$ \\
VD2 & $0.12 \pm 0.14$ & $0.26 \pm 0.10$ & $0.73 \pm 0.04$ & $0.90 \pm 0.03$ & $0.95 \pm 0.06$ \\
\bottomrule
\end{tabular}
}
\end{table}

\begin{table}[h]
\centering
\caption{Synthetic-real EKYT embedding similarity on TEX using subject-level 95\% bootstrap confidence intervals. 
$p$-values are from paired Wilcoxon signed-rank tests against DiffEyeSyn.}
\label{tab:bootstrap_test}
\resizebox{0.35\textwidth}{!}{%
\begin{tabular}{lcc}
\hline 
Model & Cosine Similarity $\uparrow$ & $p$ vs. DiffEyeSyn \\
\hline 
SP-EyeGAN  & 0.21 [0.11, 0.31] & $<0.001$ \\
VAE        & 0.73 [0.72, 0.73] & $<0.001$ \\
DiffEyeSyn & 0.94 [0.94, 0.95] & -- \\
\hline 
\end{tabular}%
}
\end{table}

\subsection{Synthetic Data Similarity}
To evaluate how closely synthetic signals preserve real eye-movement characteristics, we computed cosine similarity between EKYT \cite{lohr2022ekyt} embeddings extracted from synthetic and real sequences. Higher values (closer to 1.0) indicate stronger preservation of salient real gaze dynamics. As shown in Table~\ref{tab:cosine_similarity}, DiffEyeSyn consistently achieves the highest similarity across all tasks, outperforming both SP-EyeGAN and the auxiliary VAE baseline. For example, on TEX, where the original SP-EyeGAN was evaluated, DiffEyeSyn obtains 0.94 $\pm$ 0.06, compared with 0.73 $\pm$ 0.03 for VAE and 0.21 $\pm$ 0.10 for SP-EyeGAN. A similar trend is observed across BLG, FXS, HSS, RAN, VD1, and VD2, indicating that DiffEyeSyn better preserves subject-specific gaze dynamics across diverse tasks.

Compared with its baseline variant, DiffEyeSyn also shows consistent improvement, suggesting that the compact 128-D embedding and identity-removed conditioning provide more effective guidance for subject-specific synthesis. In contrast, SP-EyeGAN improves over its baseline but remains substantially below DiffEyeSyn, while VAE provides a stronger auxiliary baseline than SP-EyeGAN but still fails to match the diffusion-based model. To assess statistical reliability, Table~\ref{tab:bootstrap_test} reports subject-level 95\% bootstrap confidence intervals and paired Wilcoxon signed-rank tests on TEX. DiffEyeSyn achieves the highest similarity, 0.94 [0.94, 0.95], and significantly outperforms both SP-EyeGAN and VAE ($p<0.001$), confirming that its improvement is statistically consistent across subjects.

\begin{figure*}[htbp]
    \centering
    \includegraphics[width=\linewidth]{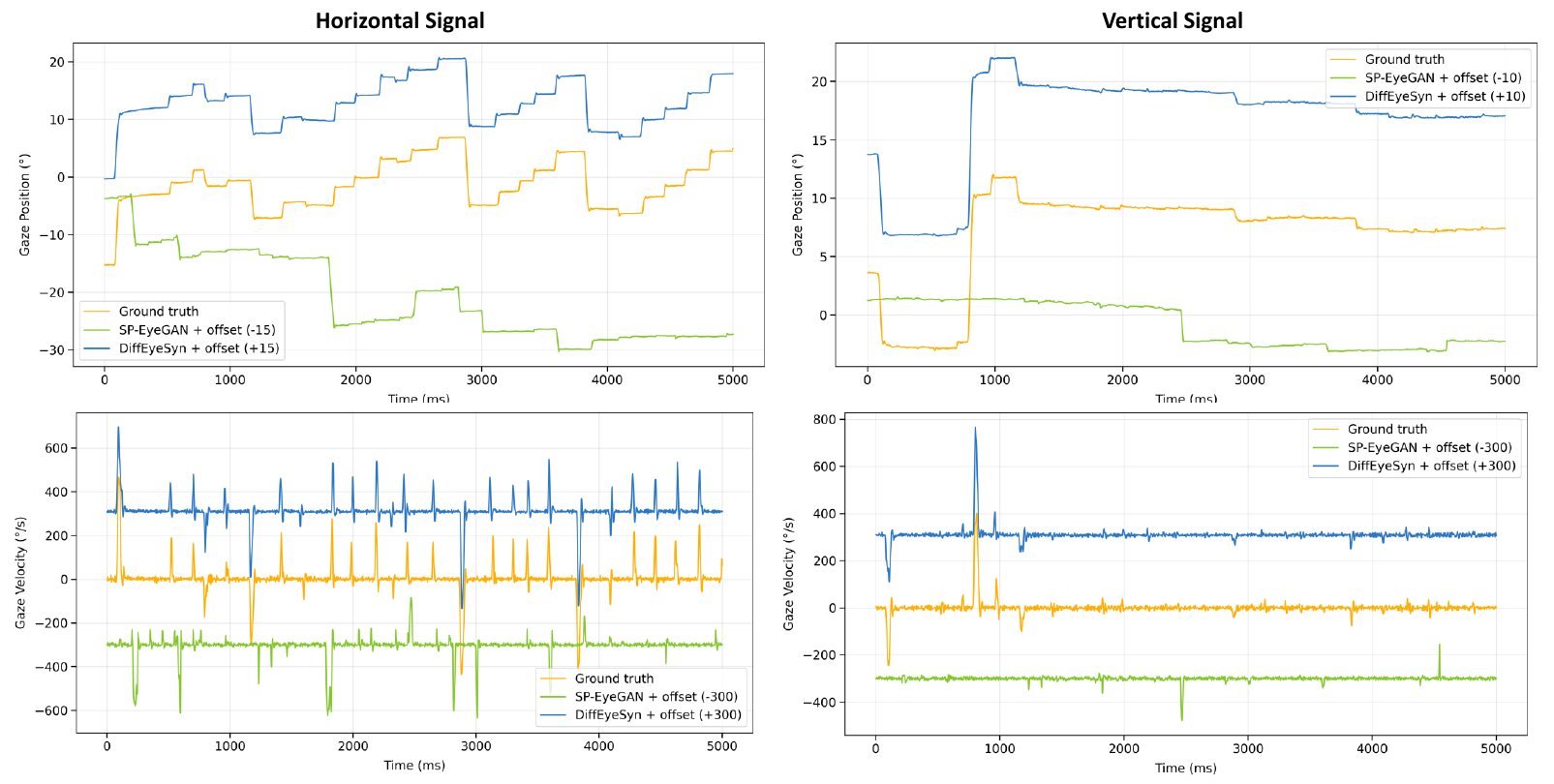}
    \caption{Qualitative comparison between SP-EyeGAN and DiffEyeSyn, where the first row contains the positional signal and the second row contains the velocity signal for horizontal and vertical eye movements.}
    \label{fig:good_generation_example}
\end{figure*}

\subsection{Qualitative Evaluations}
Figure \ref{fig:good_generation_example} compares real eye-movement signals with synthetic signals generated by SP-EyeGAN and DiffEyeSyn across two dimensions (horizontal and vertical) for both gaze position and velocity. In visual inspection, both approaches produce qualitatively realistic eye movement sequences; however, DiffEyeSyn produces more accurate, human-like signals than SP-EyeGAN. In particular, DiffEyeSyn’s synthetic sequences more closely match the ground truth patterns, exhibiting smoother velocity profiles with less jitter and better-aligned fixation and saccade segments. This improvement is due to the way DiffEyeSyn is customized: its conditional diffusion model and subject-specific, lightweight user embeddings enable it to capture subtle subjective dynamics in eye movement data, resulting in synthetic signals that are more like the real ones.

\subsection{Authentication Under Real and Synthetic Enrollment/Probe Conditions}
Table~\ref{tab:user_authentication} reports authentication performance on the TEX task under four enrollment/probe settings using the pre-trained EKYT ~\cite{lohr2022ekyt} model. Real-real matching provides the upper-bound performance, reaching 100.00\% Rank-1 IR and 0.00\% EER. In contrast, all cross-domain settings show a substantial performance drop, indicating a clear real-synthetic distribution gap. Nevertheless, DiffEyeSyn consistently outperforms SP-EyeGAN and the auxiliary VAE baseline. In the synthetic-synthetic setting, DiffEyeSyn achieves 62.71\% Rank-1 IR and 16.89\% EER, compared with 35.59\% Rank-1 IR and 24.99\% EER for VAE and 11.69\% Rank-1 IR and 41.61\% EER for SP-EyeGAN. These results indicate that DiffEyeSyn preserves more subject-discriminative structure than SP-EyeGAN and VAE, especially when both gallery and probe samples are generated synthetically. At the same time, the weak real-synthetic and synthetic-real results reveal a clear domain gap, meaning that synthetic gaze is not yet directly interchangeable with real gaze for operational authentication. From a biometric perspective, this is still an encouraging outcome, as it shows that subject-specific synthesis can retain meaningful identity cues; however, reducing the real-to-synthetic domain gap remains necessary before such data can be reliably used in practical cross-domain authentication scenarios.

\begin{table}[h]
\centering
\caption{Authentication performance under real and synthetic enrollment-probe conditions, evaluated using Rank-1 identification rate (IR) and equal error rate (EER) with the pre-trained EKYT model. Session 1 (S1) is used for enrollment and Session 2 (S2) for probe; ‘--’ denotes no generative approach is used.}
\label{tab:user_authentication}
\resizebox{\columnwidth}{!}{%
\begin{tabular}{lllcc}
\toprule
Enrollment & Probe & Model & Rank-1 IR (\%) & EER (\%) \\
\midrule
Real & Real & -- & 100.00 & 0.00 \\
\midrule
\multirow{3}{*}{Real} & \multirow{3}{*}{Synthetic}
  & SP-EyeGAN  & 1.69  & 49.80 \\
& & VAE        & 1.70  & 47.49 \\
& & DiffEyeSyn & 6.78  & 38.98 \\
\midrule
\multirow{3}{*}{Synthetic} & \multirow{3}{*}{Real}
  & SP-EyeGAN  & 1.69  & 50.85 \\
& & VAE        & 3.39  & 48.31 \\
& & DiffEyeSyn & 11.86 & 38.84 \\
\midrule
\multirow{3}{*}{Synthetic} & \multirow{3}{*}{Synthetic}
  & SP-EyeGAN  & 11.69 & 41.61 \\
& & VAE        & 35.59 & 24.99 \\
& & DiffEyeSyn & 62.71 & 16.89 \\
\bottomrule
\end{tabular}
}
\end{table}

\begin{figure}[htbp]
    \centering
    \includegraphics[width=0.8\columnwidth]{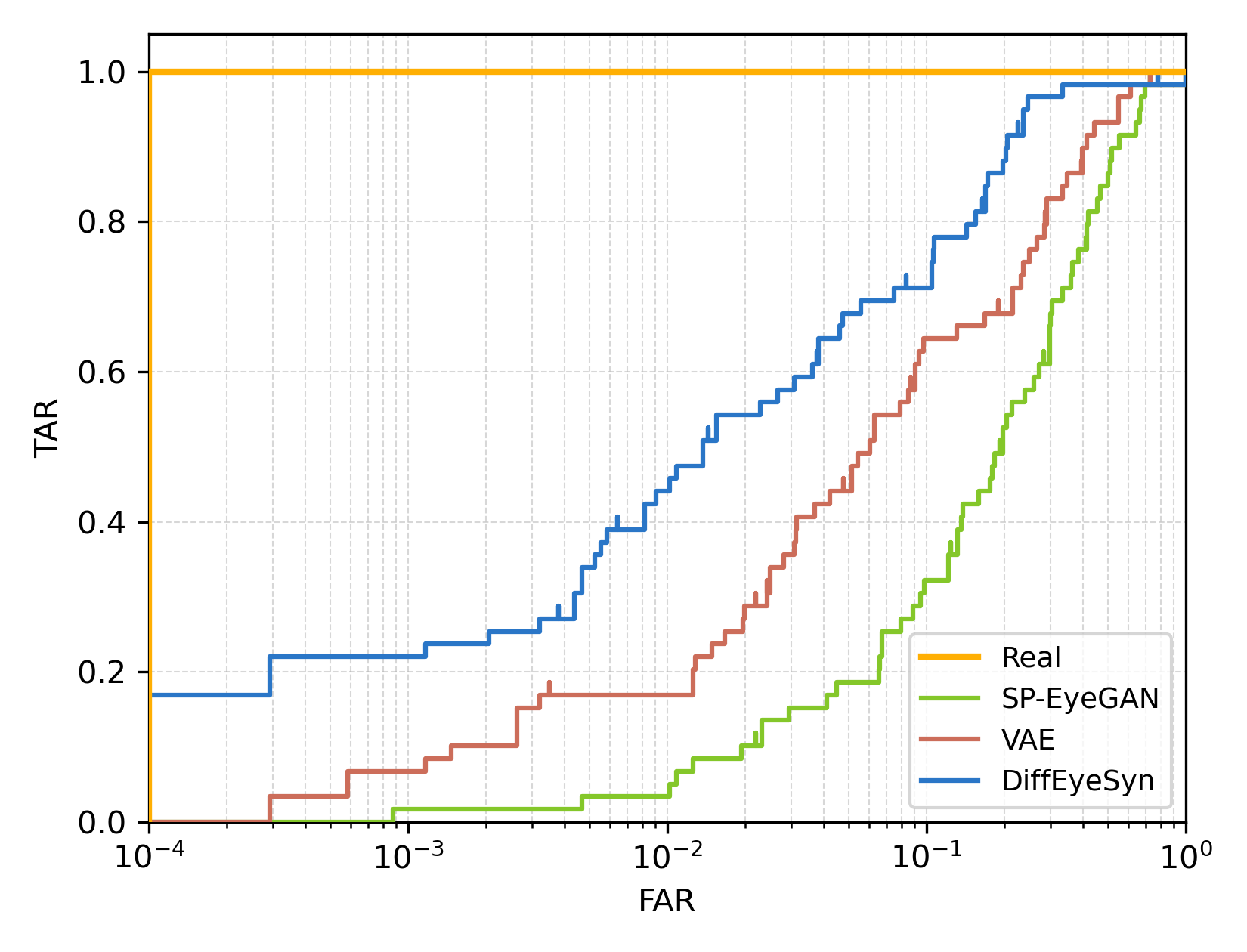}
    \caption{ROC curves for biometric verification using EKYT embeddings. FAR denotes false accept rate, and TAR denotes true accept rate; higher curves indicate stronger identity preservation.}
    \label{fig:roc_curve}
\end{figure}

To further analyze verification behavior beyond EER, Figure~\ref{fig:roc_curve} presents ROC curves using EKYT embeddings, and Table~\ref{tab:tar_far} reports TAR at fixed FAR operating points. DiffEyeSyn obtains the strongest verification performance among the synthetic methods, particularly at stricter FAR thresholds. It reaches 23.40\%, 45.43\%, and 74.19\% TAR at FAR = 0.1\%, 1\%, and 10\%, respectively, outperforming both SP-EyeGAN and VAE. The gap from real gaze remains substantial, further confirming the real-synthetic domain mismatch.

\begin{table}[h]
\centering
\small
\caption{Verification performance measured as TAR (\%) at fixed FAR operating points.}
\label{tab:tar_far}
\resizebox{\columnwidth}{!}{%
\begin{tabular}{lccc}
\toprule
Model & TAR@FAR$_{0.1\%}$ & TAR@FAR$_{1\%}$ & TAR@FAR$_{10\%}$\\
\midrule
Real       & 100.00 & 100.00 & 100.00 \\
SP-EyeGAN  & 1.75   & 5.02   & 32.34  \\
VAE        & 7.98   & 19.86  & 64.54  \\
DiffEyeSyn & 23.40  & 45.43  & 74.19  \\
\bottomrule
\end{tabular}
}
\end{table}

To verify that this identity-preservation trend is not specific to the EKYT embedding space, Table~\ref{tab:independent_authentication} reports an EKYT-independent evaluation using a ResNet1D-based authenticator \cite{saheed2024resnet50}. The same overall ranking is observed: DiffEyeSyn achieves the best synthetic-synthetic performance, with 49.15\% Rank-1 IR and 15.58\% EER, outperforming both VAE and SP-EyeGAN. This supports the conclusion that DiffEyeSyn's identity-preservation advantage is not solely an artifact of the EKYT evaluator.

\begin{table}[h]
\centering
\caption{EKYT-independent biometric validation using a ResNet1D-based authenticator.}
\label{tab:independent_authentication}
\resizebox{\columnwidth}{!}{%
\begin{tabular}{lllcc}
\toprule
Enrollment & Probe & Model & Rank-1 IR (\%) & EER (\%) \\
\midrule
Real & Real & --- & 100.00 & 0.88 \\
\midrule
\multirow{3}{*}{Real} & \multirow{3}{*}{Synthetic}
  & SP-EyeGAN  & 1.70  & 49.15 \\
& & VAE        & 1.70  & 49.15 \\
& & DiffEyeSyn & 6.78  & 42.84 \\
\midrule
\multirow{3}{*}{Synthetic} & \multirow{3}{*}{Real}
  & SP-EyeGAN  & 3.39  & 50.00 \\
& & VAE        & 3.40  & 49.86 \\
& & DiffEyeSyn & 11.86 & 39.60 \\
\midrule
\multirow{3}{*}{Synthetic} & \multirow{3}{*}{Synthetic}
  & SP-EyeGAN  & 13.56 & 33.90 \\
& & VAE        & 22.04 & 27.56 \\
& & DiffEyeSyn & 49.15 & 15.58 \\
\bottomrule
\end{tabular}
}
\end{table}

\begin{figure}[htbp]
    \centering
    \includegraphics[width=0.8\columnwidth]{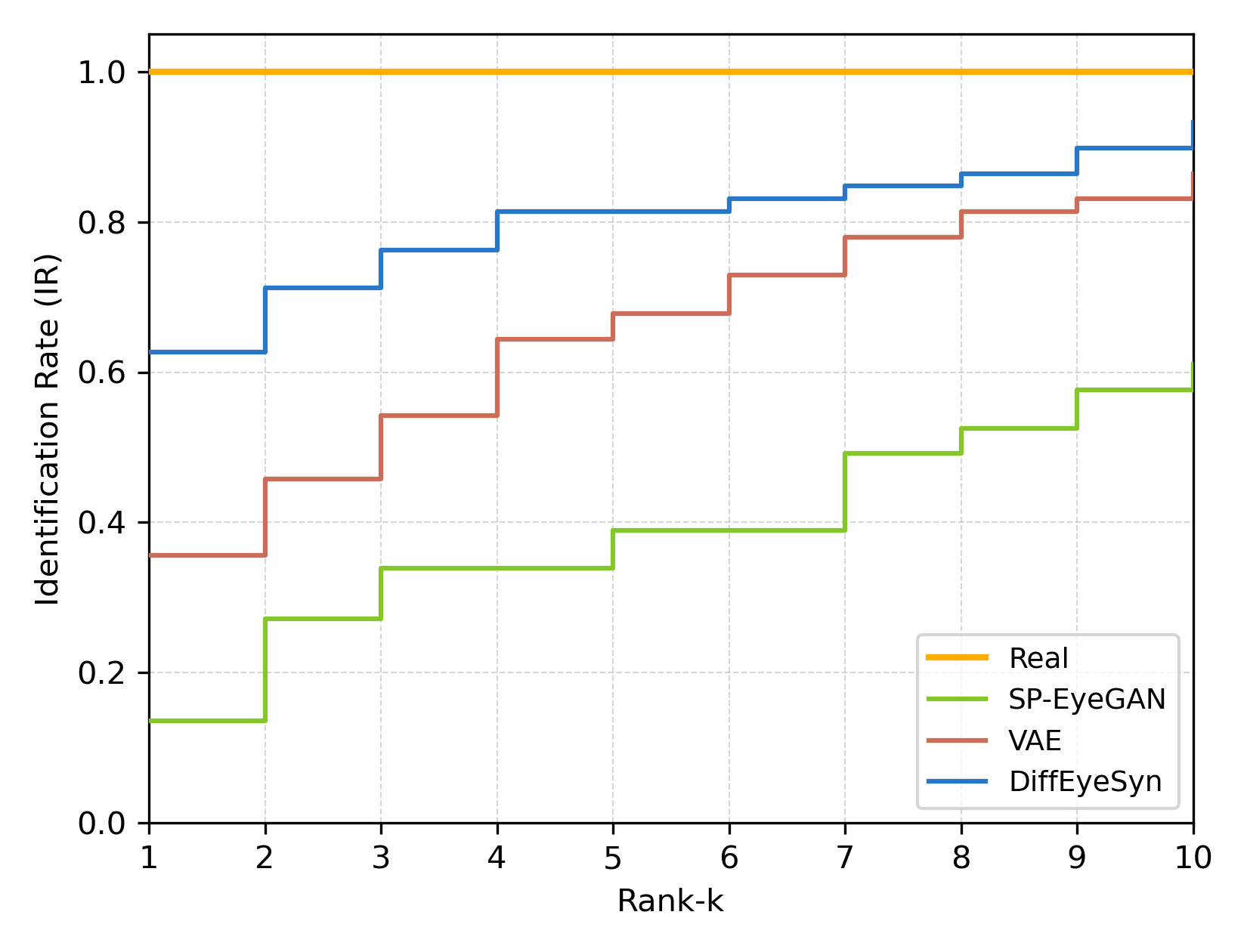}
    \caption{CMC curves for closed-set identification using EKYT embedding cosine similarities. Higher curves indicate better identity preservation.}
    \label{fig:cmc_curve}
\end{figure}

Finally, Figure~\ref{fig:cmc_curve} reports CMC curves for closed-set identification using EKYT embedding cosine similarities. DiffEyeSyn maintains the highest identification rate across Rank-$k$ values among the synthetic methods, indicating stronger preservation of subject-discriminative structure than SP-EyeGAN and VAE. This complements the Rank-1 IR results in Tables~\ref{tab:user_authentication} and~\ref{tab:independent_authentication} by showing that DiffEyeSyn remains superior beyond only the top-ranked match.

\subsection{Ablation study}
We conducted an ablation study to examine how modified features and modules affect both SP-EyeGAN and DiffEyeSyn, as presented in Table \ref{tab:ablation_study}. The Subject-Specific Condition Generator (SCG) improves SP-EyeGAN’s overall cosine similarity from 0.13 ± 0.14 to 0.24 ± 0.10. In DiffEyeSyn, the initial similarity score was 0.90 ± 0.04, and using a more lightweight conditional embedding (i.e., 128-D) reduced signal generation interference, improving the similarity score by 0.02 ± 0.06, while adding the 25 Hz identity-removed signal provides a larger improvement to 0.94 $\pm$ 0.03. However, adding both 128-D compact embeddings and a 25 Hz signal as conditions to the diffusion model yields an overall similarity score of 0.95 ± 0.06. This improvement indicates that, with fewer interruptions in the diffusion and a more detailed 25 Hz signal, these conditions can generate a more realistic subject-specific eye-movement signal.

\begin{table}[h]
\centering
\caption{Cosine-similarity ablation for SP-EyeGAN and DiffEyeSyn. \checkmark{} indicates an included component; ``--'' denotes the baseline.}
\label{tab:ablation_study}
\resizebox{0.65\columnwidth}{!}{%
\begin{tabular}{llcc}
\toprule
Model & \multicolumn{2}{c}{Component} & Overall \\
\midrule
\multirow{3}{*}{SP-EyeGAN}
  & \multicolumn{2}{c}{SCG}        &                 \\
  \cmidrule(lr){2-3}
  & \multicolumn{2}{c}{--}         & $0.13 \pm 0.14$ \\
  & \multicolumn{2}{c}{\checkmark} & $0.24 \pm 0.10$ \\
\midrule
\multirow{5}{*}{DiffEyeSyn}
  & 25\,Hz     & 128-D      &                 \\
  \cmidrule(lr){2-3}
  & --         & --         & $0.90 \pm 0.04$ \\
  & --         & \checkmark & $0.92 \pm 0.04$ \\
  & \checkmark & --         & $0.94 \pm 0.03$ \\
  & \checkmark & \checkmark & $0.95 \pm 0.06$ \\
\bottomrule
\end{tabular}
}
\end{table}

\section{Discussion}

\subsection{Interpreting Visual Quality Findings}
Generating synthetic gaze signals that fully match real ones is challenging, as realistic eye movements must preserve fine-grained characteristics such as saccades, fixations, amplitudes, and velocity patterns. As illustrated in Figure \ref{fig:good_generation_example}, DiffEyeSyn generally produces signals that more closely resemble the ground truth than SP-EyeGAN. For example, in the horizontal signal around 1000 $ms$, DiffEyeSyn follows the real pattern well, although it introduces an extra microsaccade before fixation that is not present in the ground truth. In contrast, SP-EyeGAN only partially follows the real trajectory and often fails to reproduce the correct sequence of fixations and saccades.
A similar pattern is observed in the vertical signal. SP-EyeGAN exhibits clearer mismatches, including unstable fixations, unnatural drift, incorrect saccade structure, and spurious high-velocity spikes that are absent from the real signal. DiffEyeSyn better preserves the overall temporal structure and captures some fine-grained events, but it still does not reproduce all temporal details with full fidelity. Overall, the qualitative results support the quantitative findings: DiffEyeSyn produces more realistic gaze behavior than SP-EyeGAN, but both approaches remain imperfect approximations of real eye movements.

\subsection{Synthetic Signal Quality and Subject Specificity}
The results in Tables~\ref{tab:spatial_accuracy}, \ref{tab:spatial_precision}, and~\ref{tab:cosine_similarity} show that DiffEyeSyn produces more accurate, stable, and subject-consistent synthetic gaze signals than SP-EyeGAN. This improvement is driven by the combination of 25 Hz identity-removed guidance and compact 128-D subject embedding, which provides effective conditioning for subject-specific synthesis. Although the modified SP-EyeGAN benefits from DDQFE and OH conditioning, it remains behind DiffEyeSyn in temporal fidelity and synthetic-real similarity. The user- and error-percentile analysis further shows that DiffEyeSyn maintains lower median spatial error under typical conditions, achieving 3.73 dva for HSS and 4.06 dva for RAN at U50\textbar E50, compared with 15.45 and 13.63 dva for SP-EyeGAN, indicating more reliable gaze synthesis across users and conditions.

\subsection{Biometric Utility and Real--Synthetic Domain Gap}
DiffEyeSyn preserves more identity-discriminative information than SP-EyeGAN and the auxiliary VAE baseline, with consistent improvements across EKYT-based authentication, ROC/TAR analysis, CMC identification, and EKYT-independent ResNet1D validation. However, weak real-synthetic and synthetic-real performance indicates a remaining domain gap, meaning that synthetic gaze is not yet interchangeable with real gaze for operational authentication. Reducing this mismatch remains an important direction for future work.

\subsection{Limitations and Future Directions}
This study focuses on high-frequency, lab-grade gaze signals and may not directly generalize to lower-fidelity consumer eye trackers. Both SP-EyeGAN and DiffEyeSyn also rely on substantial training data, preprocessing, and subject-specific conditioning. Future work should explore scalable conditioning strategies, such as self-supervised embeddings and task or stimulus cues, to improve generalization and cross-domain biometric utility.

\section{Conclusion}
In this work, we adapted SP-EyeGAN and DiffEyeSyn for subject-specific gaze synthesis and compared them using spatial accuracy, spatial precision, synthetic-real similarity, qualitative inspection, and biometric verification/identification metrics. DiffEyeSyn consistently outperformed SP-EyeGAN, producing more realistic gaze signals and better preserving subject-specific dynamics. However, cross-domain authentication results show that synthetic gaze is not yet directly interchangeable with real gaze for operational biometric use. Overall, diffusion-based subject-specific synthesis appears to be a promising direction for eye-movement biometrics, where signal fidelity and identity preservation must be jointly maintained.

{\small
\bibliographystyle{ieee}
\bibliography{egbib}
}

\end{document}